\documentclass{svproc}
\usepackage{graphicx}

\usepackage[usenames,dvipsnames,table]{xcolor}
\usepackage{url}

\begin{document}

\newcommand{\maria}[1]{{\textcolor{blue!80!black}{Maria: #1}}}

\title{Visualisation for the CIS benchmark scanning results}

\author{Zhenshuo Zhao, Maria Spichkova, Duttkumari Champavat, Juilee N. Kulkarni, Sahil Singla, Muhammad A. Zulkefli, Pradhuman Khandelwal}
\authorrunning{Z. Zhao et al.}
%
\institute{School of Computing Technologies, RMIT University, Australia 
}
\maketitle              
\begin{abstract} 
In this paper, we introduce GraphSecure, a web application that provides advanced analysis and visualisation of security scanning results. GraphSecure enables users to initiate
scans for their AWS account, validate them against specific Center for Internet Security (CIS)  Benchmarks and return results, showcase those returned results in the form of statistical charts and warn the users about their account status.  \\
\emph{Preprint. Accepted to the ICICT'26. Final version to be published by in conference proceedings by Springer LNNS.}

\end{abstract}
\section{Introduction}

Software systems are becoming more exposed to cyber-attacks as we increase our use of
technology. Since the software industry is moving towards cloud-based technologies
such as AWS and GCP, it is essential to have checks for vulnerabilities. 
One of the globally recognised cybersecurity approaches for vulnerability analysis is an analysis based on the \emph{CIS benchmarks}, specified by the Center for Internet Security (CIS, see~\cite{CIS}). CIS is a non-profit organisation whose aim is to improve internet security by providing a set of benchmarks and software solutions. The CIS benchmark guidelines have been developed with a global community of security experts, with the goal of proactively safeguarding against emerging risks and limiting configuration-based security vulnerabilities in software.  

The CIS benchmarks have been specified for several Cloud platforms and services, including Alibaba Cloud, Amazon Web Services, Google Cloud, and Microsoft Azure. They encompass comprehensive best-practice recommendations, each of which includes description, rationale, impact, audit steps and remediation. Despite the clear benefits of the CIS benchmarks, organisations often face significant practical challenges in their implementation. The sheer volume of controls to be enforced can be overwhelming, rendering manual compliance processes time-consuming, resource-intensive, and economically inefficient. These difficulties are further compounded by the need for continuous monitoring across a wide range of security domains, as well as the requirement to remain up to date with newly released benchmark versions that address emerging threats. Such updates not only demand additional implementation effort but also require ongoing technical upskilling of staff. Consequently, there is a clear need for automated mechanisms that can efficiently assess and visualise the security posture of cloud environments.

\textbf{Contributions:} To analyse the recent works related to the CIS benchmark adoption, we conducted a systematic review of related works.  We also elaborated a GraphSecure application that can (1) run on multiple AWS accounts, (2) analyse how secure the AWS accounts are by validating them against multiple CIS benchmarks, and (3) provide recommendations to overcome the identified issues so that no
security threat is eliminated. GraphSecure provides an automated approach to identifying and interpreting security risks within multiple AWS accounts, thereby reducing maintenance overhead and lowering the technical barrier to effective security assessment. The proposed approach focuses on visualising identified issues to assist practitioners in making informed decisions and implementing effective solutions in their respective fields.

\section{Related work}
\label{sec:related}

Spichkova et al.~\cite{spichkova2020gosecure}  
presented GoSecure, a security inspection tool that can scan multiple Google Cloud Platform (GCP) instances against industry-recognised CIS benchmarks. 
The tool addresses all categories outlined in the CIS benchmarks for GCP and offers a comprehensive overview of the security profiles for each GCP project.
The tool also provides practical suggestions to improve configurations for the individual projects. In contrast to GoSecure, the solution we present in this paper focuses on the AWS platform and also has a greater emphasis on the visualisation aspects of the security profile. 
Spichkova et al. also elaborated \emph{Virtual Machine Vending Machine} (VM2), see \cite{spichkova2020vm2}. This tool provides functionality to create virtual machine images and to test them against security benchmarks. VM2 offers a simple way to share secure images.

George et al.~\cite{george2020usage} introduced a 
novel solution for centralised management of multiple AWS  accounts. This solution enables visualisation of AWS service usage across customer accounts. 
This solution also enables cost and performance optimisation for the AWS services across the customer’s accounts.   

While our work also focuses of AWS solutions, in contrast to the above tools, our solution is focused on the visualisation of CIS benchmark scanning results.

There are some effective tools, such as Trusted Advisor \cite{TrustedAdvisor}, Config \cite{Config} and GuardDuty \cite{GuardDuty}, provided by AWS, which solve the problem. However, they work only with a single AWS account. We propose an optimisation of the solution.  
To analyse the recent works related to the CIS-based approaches, we conducted a systematic literature review (SLR) in Scopus and Google Scholar databases, following the well-established guidelines for Systematic Literature Reviews\footnote{Cited 11,783 times according to Google Scholar, as per 1 December 2025.} proposed by Kitchenham et al.~\cite{kitchenham2004procedures}. 

\textbf{Phase 1 (Automated search):} In this phase, first of all, we specified the following  search strings to conduct an automated search: 
{\small{\textit{(`CIS' and `security') }}} ~and~
{\small{\textit{(`CIS benchmark'). }}}

\textbf{Phase 2 (Filtering):} We filtered the set of publications we obtained as the result of Phase 1.   
Our \emph{inclusion criteria} for filtering has been specified as follows:
\begin{enumerate}
    \item 
    Publications written in English.
    \item 
    Publications that present research related to the CIS benchmark application or analysis. 
\end{enumerate}
Our \emph{exclusion criteria} have been specified as follows:
\begin{enumerate}
    \item 
    Publications that were not written in English.  
    \item 
    Papers that don't cover any aspects related to CIS benchmark application or analysis.
    \item 
    Duplicate publications (in the case duplicate papers have been identified, we include only the most complete, recent, and improved version of the publication).
    \item 
    Prefaces, interviews, reviews, posters, panel discussions, tutorial summaries, and article summaries, combined conference proceedings. 
    \item 
    Papers that aren't accessible online. 
\end{enumerate}

\begin{figure}[ht!]
\begin{center}
\includegraphics[width=0.85\textwidth]{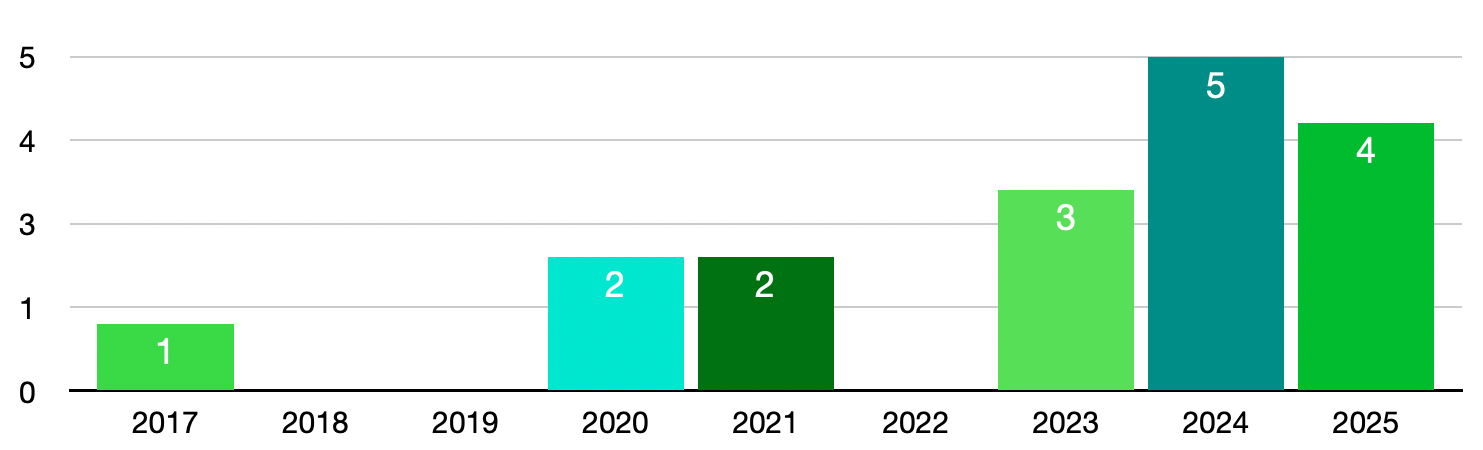}
\end{center}
\caption{Distribution of the identified studies by the year of publication} 
\label{fig:statistics}
\end{figure}

\textbf{Phase 3 (Data extraction and synthesis):} The first author extracted data from the publications, which were in the set of selected papers we obtained as a result of Phase 2. 
After this step, the second author reviewed the results in several online meetings to finalise the outcome. Overall, our search resulted in 17 relevant publications. While we haven't restricted our search by any time frame, the relevant studies have been identified as published between 2017 and 2025, with the majority of the papers published in the last three years, see Figure~\ref{fig:statistics}.  These observations highlight the growing interest in this research area. 
The summary of the analysis is presented in Table~\ref{tab:summary}. 

\begin{table}[ht!]
\centering
\caption{Summary of CIS-based studies and contributions}
\label{tab:summary}
\footnotesize
\rowcolors{2}{gray!10}{white}
\begin{tabular}{|l|l|p{7cm}|}
\hline
\textbf{Ref.} & \textbf{Type of the study } & \textbf{Study/Contribution} \\
\hline
\cite{irawan2025evaluating} & Analysis of CIS benchmark & Evaluation of the effectiveness of the CIS benchmark. \\
\cite{volotovskyi2024automated} & CIS-based analysis 
& Automating security analysis and assessments of AWS accounts through Python 3. \\
\cite{battula2023security} & CIS-based assessments & Security measures in hybrid IT infrastructures that blend on-premise systems with private and public cloud platforms. \\
\cite{Shamim20252432} & CIS-based analysis 
& Empirical study to evaluate 53 existing system configurations wrt. CIS benchmarks. Survey with 20 practitioners. \\
\cite{bar2023towards} 
& CIS-based framework & Analysis of the cybersecurity posture. \\
\cite{cue2024cis} 
& CIS-based framework & Analysis of the cybersecurity posture. \\
\cite{guduru2020cloud} & CIS-based framework & Automation of the policy-as-code enforcement workflow. \\
\cite{khurat2021automatic} & CIS-based framework & Automation of the CIS-compliance using Ansible, focusing on network devices. \\
\cite{Silva20241} & CIS-based framework & Compliance framework for IaaS cloud environments, based on CIS Controls and NIST Cybersecurity Framework recommendations. \\
\cite{sujatha2024system} & CIS-based framework & Automation of the security audits on Windows servers using Ansible. \\
\cite{xu2023optimizing} & CIS-based framework 
& Optimisation of the analysis process of the potential cybersecurity gaps, focused on legacy railway control systems. \\
\cite{Sondhiya2025581} & CIS-based framework
& Automation of the CIS benchmark compliance using AWS Config and Ansible. \\
\cite{sholihin2025oscat} & CIS-based tool & CIS-Benchmark auditing tool (OSCAT). \\
\cite{torkura2021continuous} & CIS-based tool & Automation of the CIS-compliance. \\
\cite{Edwards20241} & guidelines & CIS Critical Security Controls as a prioritized, risk-based cybersecurity defense framework. \\
\cite{anisetti2017security} & New/refined benchmark & A refinement of the CIS benchmark. \\
\cite{berkovich2020ubcis} & New/refined benchmark & Novel benchmarks for container image scanning. \\
\hline
\end{tabular} 
\end{table}

The studies summarised in Table~\ref{tab:summary}  can be broadly categorised into three main directions: (1) CIS-based security assessments, (2) CIS compliance automation frameworks and tools, and (3) the development of new or refined CIS benchmarks. CIS-based assessment studies primarily focus on evaluating the security posture of cloud environments, hybrid IT infrastructures, and existing system configurations against CIS recommendations, with some works incorporating empirical evaluations and practitioner surveys.

A substantial portion of the publications focuses on automating CIS compliance and security auditing. These solutions aim to reduce manual effort through continuous enforcement and automated discovery of cybersecurity gaps across diverse platforms, including cloud services, Windows servers, network devices, and legacy industrial systems. Ansible is frequently adopted as the primary automation tool in these frameworks, highlighting the emphasis on scalable and repeatable compliance enforcement. A smaller number of studies contribute by refining existing CIS benchmarks or proposing new benchmarks for emerging technologies, such as container image scanning.

\begin{figure}[ht!]
\begin{center}
\includegraphics[width=\textwidth]{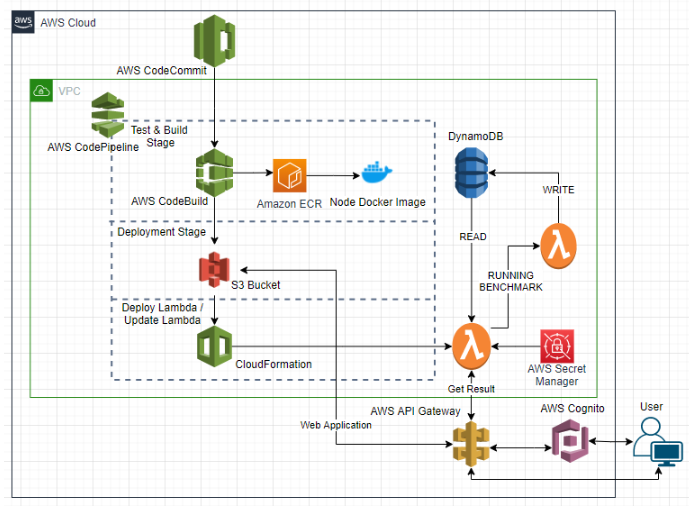}
\end{center}
\caption{System architecture of GraphSecure} 
\label{fig:arc1}
\end{figure}

 Overall, the results of our systematic literature review indicate a strong industry-driven shift toward automating CIS-based security assessments and enforcing continuous compliance in response to the increasing complexity and volume of CIS recommendations. 
However, while existing works focus heavily on auditing and enforcement, limited attention has been given to intuitive, risk-oriented visualisation of CIS compliance outcomes for cloud users. This gap motivates the development of GraphSecure, which complements existing automation approaches by providing an automated, visual representation of security risks in AWS environments to support more effective interpretation and decision-making.

\section{Proposed solution}
\label{sec:solution}  

To implement the core ideas of   GraphSecure, a serverless approach was applied, where the application is based on a combination of multiple AWS services. Figure~\ref{fig:arc1} presents the architecture of the proposed solution, while the core process is discussed below.

\emph{Step 1:} The developer makes changes to the existing code while development commits those changes to the \emph{AWS CodeCommit} repository, which runs the CI/CD pipeline.

\emph{Step 2:} \emph{CodeBuild}  bundles all the files of developed code together %
and prepares all the artifacts for placing into \emph{AWS S3}.

\emph{Step 3:} In the deployment stage, \emph{AWS S3} contains artifacts that consist of frontend code
developed in ReactJS (TypeScript), %
JavaScript code for executing CIS benchmarks and writing results to DynamoDB. 

\emph{Step 4:} %
\emph{CloudFormation} allocates associated AWS services and resources to the application.

\emph{Step 5:} Finally, the user can log in to the application using the URL provided by the API gateway, wherein various endpoints are managed for carrying functionalities like hosting front-end, executing Lambda to run scans and lastly to get results from the database. In this case, the results demonstrate that some benchmarks have failed, GraphSecure would provide recommendations with step-by-step instructions to mitigate the issue. 

The basic requirement to run this application is the installation of NodeJS v15.9.0
and Git CLI.
The development environment adheres to the following technology specifications:
React v16.12.0 or above,  TypeScript v4.2.3 or above,  Webpack v5.37.1 or above, Jest v26.6.3 or above, Chai v4.3.4 or above.


\begin{figure}[ht!]
\begin{center}
\includegraphics[width=\textwidth]{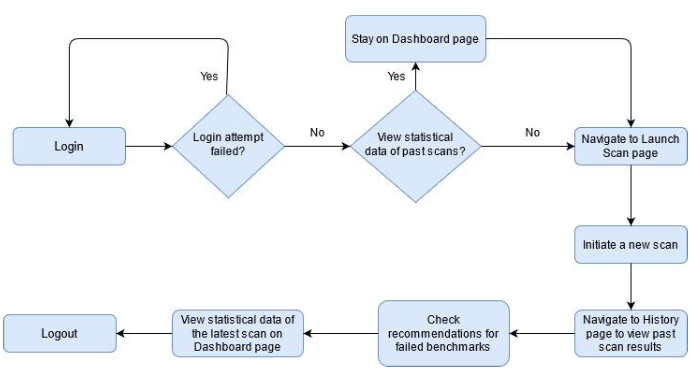}
\end{center}
\caption{High-level workflow of GraphSecure} 
\label{fig:screens1}
\end{figure}

\begin{figure}[ht!]
\begin{center}
\includegraphics[width=0.9\textwidth]{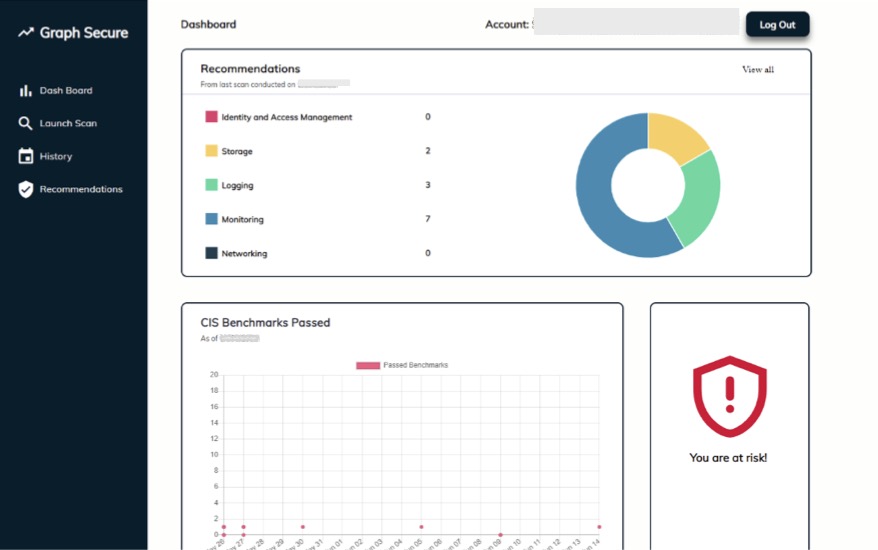}
\end{center}
\caption{GraphSecure: `Dashboard' page} 
\label{fig:GS-dashboard}
\end{figure}

\begin{figure}[ht!]
\begin{center}
\includegraphics[width=0.9\textwidth]{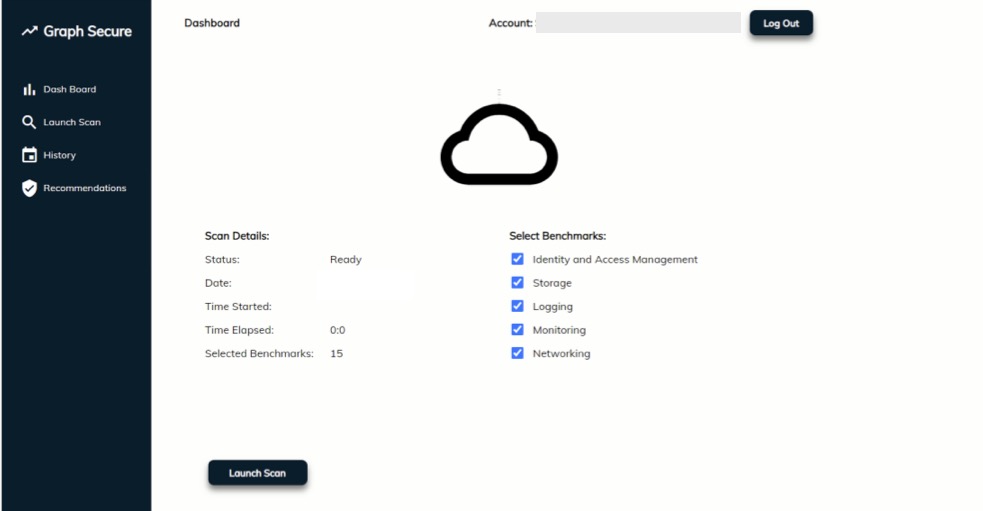} 
\end{center}
\caption{GraphSecure: `Launch Scan' page. The tool is ready to conduct a new scan.} 
\label{fig:GS-scan}
\end{figure}

Figure~\ref{fig:screens1} presents the high-level workflow. 
Once the user is successfully authenticated, they are navigated to the `Dashboard' page (see Figure~\ref{fig:GS-dashboard}), which presents 
chart components to provide visual representations of scan results.
The doughnut chart displays the recommendations suggested in the latest scan, with a category-wise count depicting the failed benchmarks of specific CIS categories (e.g., monitoring, networking, storage, logging).

To run a scan for their AWS account against CIS
benchmarks belonging to categories IAM, monitoring, networking, storage, and logging,  
the user has to navigate to the `Launch Scan' page (see Figure~\ref{fig:GS-scan}).
On that page, the user can select/deselect categories to run a scan for their AWS account against CIS
benchmarks. Once the user selects the checkboxes of their choice and clicks `Launch Scan'
button, the timer starts, and the status changes to \emph{Scanning}.

\begin{figure}[ht!]
\begin{center}
\includegraphics[width=0.85\textwidth]{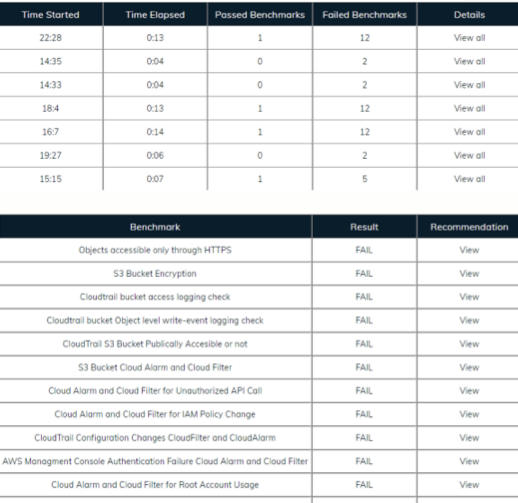}  
\end{center}
\caption{GraphSecure: Extract from the `History' page focused on the  table data} 
\label{fig:GS-history}
\end{figure}

The results of previously completed scans are presented on the  `History' page (see Figure~\ref{fig:GS-history}). 
presents results in a tabular format. 
The scan results are sorted by date and time when the scans were conducted. The other information included in the history covers the scan start time and the number of benchmarks to be checked against the selected categories. 
For each scan, the user can also see a count of passed and failed benchmarks.

To view detailed information about failed benchmarks, the user can click the `View All' button, which opens a tabular view of failed benchmarks with their names, date, and status. 
The `View' button associated with a failed benchmark directs the user to the `Recommendations' page, where the recommendations on how to rectify the required security risk step-by-step are provided.

This support would help solve the issue by following the clear and straightforward steps, ensuring that even novice developers can easily navigate and utilise the tool. Thus, while the visualisation functionality of GraphSecure might be useful for professionals regardless of their practical security experience, the recommendations-related functionality would be especially helpful for novice developers.

\section{Discussion and Conclusions} 
\label{sec:conclusions}

In this paper, we presented a novel tool GraphSecure. The tool provides an automated analysis of how secure the AWS accounts are by checking them against multiple CIS
benchmarks. The tool also provides recommendations to overcome failed benchmarks. One of the advantages of the tool is that it can be applied to multiple AWS accounts. It provides easy-to-read visualisation of the scan results to help identify and interpret security risks. 
Compared to the approaches presented in Section~\ref{sec:related}, GraphSecure provides functionality similar to the solutions introduced in \cite{Sondhiya2025581} and \cite{volotovskyi2024automated}. However, compared to the above tools, GraphSecure provides more emphasis on the practical usability of the proposed solution, aiming to increase the readability of the presented information.  

\section*{{Acknowledgements}}
 
We would like to thank Shine Solutions Group Pty Ltd for supporting this project under a research grant PRJ00000343.

\bibliographystyle{splncs04}

\begin{thebibliography}{10}
\providecommand{\url}[1]{\texttt{#1}}
\providecommand{\urlprefix}{URL }
\providecommand{\doi}[1]{https://doi.org/#1}

\bibitem{anisetti2017security}
Anisetti, M., Ardagna, C.A., Damiani, E., Gaudenzi, F.: A security benchmark for openstack. In: 2017 IEEE 10th International Conference on Cloud Computing (CLOUD). pp. 294--301. IEEE (2017)

\bibitem{GuardDuty}
{AWS}: {Amazon GuardDuty}. \url{https://aws.amazon.com/guardduty/} (2025), online, last accessed 28 November 2025.

\bibitem{Config}
{AWS}: {AWS Config}. \url{https://aws.amazon.com/config/} (2025), online, last accessed 28 November 2025.

\bibitem{TrustedAdvisor}
{AWS}: {AWS Trusted Advisor}. \url{https://aws.amazon.com/premiumsupport/technology/trusted-advisor/} (2025), online, last accessed 28 November 2025.

\bibitem{bar2023towards}
Bar-Haim, R., Eden, L., Kantor, Y., Agarwal, V., Devereux, M., Gupta, N., Kumar, A., Orbach, M., Zan, M.: Towards automated assessment of organizational cybersecurity posture in cloud. In: Proc. of the 6th Joint International Conference on Data Science \& Management of Data. pp. 167--175 (2023)

\bibitem{battula2023security}
Battula, V.: Security compliance in hybrid environments using tripwire and cyberark. International Journal of Research and Analytical Reviews  \textbf{10}(2),  788--803 (2023)

\bibitem{berkovich2020ubcis}
Berkovich, S., Kam, J., Wurster, G.: {UBCIS}: Ultimate benchmark for container image scanning. In: 13th USENIX Workshop on Cyber Security Experimentation and Test (CSET 20) (2020)

\bibitem{CIS}
{Center for Internet Security}: {CIS Benchmarks List}. \url{https://www.cisecurity.org/cis-benchmarks} (2025), online, last accessed 28 November 2025.

\bibitem{cue2024cis}
Cue, H.A.A., Bourlai, T., Lupo, M.: A cis controls v8. 0 scoring system using combined ranking-weight methods. In: 2024 IEEE International Systems Conference (SysCon). pp.~1--8. IEEE (2024)

\bibitem{Edwards20241}
Edwards, D.J.: Critical Security Controls for Effective Cyber Defense. Springer (2024)

\bibitem{george2020usage}
George, L.C., Guo, Y., Stepanov, D., Peri, V.K.R., Elvitigala, R.L., Spichkova, M.: Usage visualisation for the {AWS} services. Procedia Computer Science  \textbf{176},  3710--3717 (2020)

\bibitem{guduru2020cloud}
Guduru, S.: {Cloud Security Automation: Enforcing CIS Benchmarks with AWS Config, Azure Policy, and OpenStack Chef Cookbooks}. Journal of Scientific and Engineering Research  \textbf{7}(10),  243--248 (2020)

\bibitem{irawan2025evaluating}
Irawan, B., Sheha, K.N., Rahaman, M., Erzed, N., Herwanto, A.: Evaluating the effectiveness of center of internet security benchmark for hardening linux servers against cyber attacks. Journal of Social Research  \textbf{4}(6),  1172--1183 (2025)

\bibitem{khurat2021automatic}
Khurat, A., Sangkhachantharanan, P.: An automatic networking device auditing tool based on cis benchmark. In: 2021 18th International Conference on Electrical Engineering/Electronics, Computer, Telecommunications and Information Technology (ECTI-CON). pp. 409--412. IEEE (2021)

\bibitem{kitchenham2004procedures}
Kitchenham, B.: Procedures for performing systematic reviews. Keele, UK, Keele University  \textbf{33}(2004),  1--26 (2004)

\bibitem{Shamim20252432}
Shamim, S.I., Hu, H., Rahman, A.: On prescription or off prescription? {An} empirical study of community-prescribed security configurations for {Kubernetes}. International Conference on Software Engineering p. 2432 – 2444 (2025)

\bibitem{sholihin2025oscat}
Sholihin, A., Salman, M.: Oscat: A comprehensive tool for automated cis benchmark auditing. Asian Journal of Engineering, Social and Health  \textbf{4}(2),  443--452 (2025)

\bibitem{Silva20241}
Silva, R., Brito, A., de~Lima~Filho, J.P.: Automation of security controls for continuous compliance in vulnerability management. In: Proceedings of the 13th Latin-American Symposium on Dependable and Secure Computing. pp. 1--10 (2024)

\bibitem{Sondhiya2025581}
Sondhiya, L., Pathi, N.K., Agarwal, R.: A framework for automating compliance as code using {AWS Config and Ansible}. Smart Innovation, Systems and Technologies p. 581 – 593 (2025)

\bibitem{spichkova2020vm2}
Spichkova, M., Li, B., Porter, L., Mason, L., Lyu, Y., Weng, Y.: {VM2:} automated security configuration and testing of virtual machine images. Procedia Computer Science  \textbf{176},  3610--3617 (2020)

\bibitem{spichkova2020gosecure}
Spichkova, M., Vaish, A., Highet, D.C., Irfan, I., Kesley, K., Kumar, P.D.: {GoSecure: Securing Projects with Go.} In: ENASE. pp. 587--594 (2020)

\bibitem{sujatha2024system}
Sujatha, G., et~al.: System hardening using cis benchmarks. In: 2024 International Conference on Advances in Computing, Communication and Applied Informatics (ACCAI). pp.~1--6. IEEE (2024)

\bibitem{torkura2021continuous}
Torkura, K.A., Sukmana, M.I., Cheng, F., Meinel, C.: Continuous auditing and threat detection in multi-cloud infrastructure. Computers \& Security  \textbf{102},  102124 (2021)

\bibitem{volotovskyi2024automated}
Volotovskyi, O., Banakh, R., Piskozub, A., Brzhevska, Z.: {Automated security assessment of Amazon Web Services accounts using CIS Benchmark and Python 3}. Cybersecurity Providing in Information and Telecommunication Systems II 2024  \textbf{3826},  363--371 (2024)

\bibitem{xu2023optimizing}
Xu, W.: Optimizing cyber security gap analysis for legacy railway control systems: A proposed new gap analysis process using cis benchmarks™ (2023)

\end{thebibliography}

\end{document}